\documentclass[copyright,creativecommons]{eptcs}
\providecommand{\event}{TERMGRAPH 2011}
\usepackage{graphicx}
\usepackage{mathrsfs}
\usepackage{amssymb}
\usepackage{stmaryrd}
\usepackage[T1]{fontenc}
\usepackage[utf8]{inputenc}
\usepackage{subfigure}
\usepackage{url}
\usepackage{color}

\usepackage{footmisc}

\usepackage{tabularx}
\usepackage{longtable}
\urldef{\porgydl}\url{http://gravite.labri.fr/?Projects_%2F_Grants:Porgy:Download}
\urldef{\gravite}\url{http://gravite.labri.fr/?Projects_%2F_Grants:Porgy}

{}

\newcommand{\sema}[1] {\ensuremath{\left [ #1\right ]}\xspace} 

\newcommand{\ident}{{\tt Id}}
\newcommand{\Fail}{{\tt Fail}}

\newcommand{\ra}{\rightarrow}
\newcommand{\Ra}{\Rightarrow}

\newcommand{\R}{{\cal R}}

\newcommand{\X}{{\cal X}}

\newcommand{\Interface}{\mathit{Interface}}

\ifx\thm\undefined
\newtheorem{thm}{Theorem}
\fi

\ifx\exmp\undefined
\newtheorem{exmp}[thm]{Example}
\fi

\ifx\definition\undefined
\newtheorem{definition}[thm]{Definition}
\fi

\newcommand{\AKAP}{{\sf AKAP}}
 \newcommand{\cAMP}{{\sf cAMP}}
 \newcommand{\MEK}{{\sf MEK}}
 \newcommand{\ERK}{{\sf ERK}}
 
 \newcommand{\PKA}{{\sf PKA}}
 \newcommand{\Raf}{{\sf Raf-1}}

 \newcommand{\PDE}{{\sf PDE8}}
 
 \newcommand{\SA}{{\sf SA}}

\sloppy

\usepackage{xspace}

\newcommand\putnline[3]
{\count0 = #2 \advance \count0 by 10%
\put(#1,#2){\line(0,1){#3}}%
\put(#1,\count0){\makebox(0,0){$/$}}}

\newcommand\putaxiom[3]
{\count2 = #3 \count0 = #3 \divide \count0 by 2
\count1 = #1 \advance \count1 by \count0
\put(\count1,#2){\line(-1,0){#3}}%
\put(\count1,#2){\line(0,-1){10}}%
\advance \count1 by -\count2%
\put(\count1,#2){\line(0,-1){10}}%
}

\newcommand\putcut[3]
{\count2 = #3 \count0 = #3 \divide \count0 by 2
\count1 = #1 \advance \count1 by \count0
\put(\count1,#2){\line(-1,0){#3}}%
\put(\count1,#2){\line(0,1){10}}%
\advance \count1 by -\count2%
\put(\count1,#2){\line(0,1){10}}%
}

\newcommand\putdtriangle[3]{\count0=#2 \advance \count0 by 9%
\count1=#1 \advance \count1 by -10%
\count2=#2 \advance \count2 by 15%
\put(#1,#2){\line(-2,3){10}}%
\put(#1,#2){\line(2,3){10}}%
\put(\count1,\count2){\line(1,0){20}}%
\advance \count1 by 4%
\put(\count1,\count2){\line(0,1){10}}%
\advance \count1 by 12%
\put(\count1,\count2){\line(0,1){10}}%
\put(#1,#2){\vector(0,-1){10}}%
\put(#1,\count0){\makebox(0,0){#3}}}

\newcommand{\footnoteremember}[2]{
  \footnote{#2}
  \newcounter{#1}
  \setcounter{#1}{\value{footnote}}
}
\newcommand{\footnoterecall}[1]{
  \footnotemark[\value{#1}]
}

\title{PORGY: Strategy-Driven Interactive Transformation\\
  of Graphs\thanks{Partially supported by INRIA's
    \emph{Associate team program} (see \gravite) and the French National
    Research Agency project EVIDEN (ANR 2010 JCJC 0201 01).}}

\author{Oana Andrei\footnote{School of Computing Science, University
    of Glasgow, Glasgow G12 8RZ, UK}\email{oana.andrei@glasgow.ac.uk}
\and
Maribel Fern\'andez\footnoteremember{kings}{King's College London,
  Department of Informatics, Strand, London WC2R 2LS,
  UK}\email{maribel.fernandez@kcl.ac.uk}
\and
H\'el\`ene Kirchner\footnoteremember{inria-labri}{INRIA Bordeaux
  Sud-Ouest, Universit\'e Bordeaux 1, CNRS UMR 5800, LaBRI, 33405
  Talence Cedex, France}\email{helene.kirchner@inria.fr}
\and
Guy Melan\c{c}on\footnoterecall{inria-labri}\email{guy.melancon@inria.fr}
\and
Olivier Namet\footnoterecall{kings}\email{olivier.namet@kcl.ac.uk}
\and
Bruno Pinaud\footnoterecall{inria-labri}\email{bruno.pinaud@inria.fr}
}

\def\titlerunning{PORGY}
\def\authorrunning{Andrei \emph{et al.}}

\begin{document}
\maketitle

\begin{abstract}
  This paper investigates the use of graph rewriting systems as a
  modelling tool, and advocates the embedding of such systems in an
  interactive environment.  One important application domain is the
  modelling of biochemical systems, where states are represented by
  port graphs and the dynamics is driven by rules and strategies.  A
  graph rewriting tool's capability to interactively explore the
  features of the rewriting system provides useful insights into
  possible behaviours of the model and its properties.  We describe
  PORGY, a visual and interactive tool we have developed to model
  complex systems using port graphs and port graph rewrite rules
  guided by strategies, and to navigate in the derivation history.  We
  demonstrate via examples some functionalities provided by PORGY.
  \end{abstract}

\section{Introduction} %

Graphical formalisms are widely used for describing complex structures
in a visual and intuitive way, \emph{e.g.}, UML diagrams,
representation of proofs, microprocessor design, XML documents,
communication networks, data and control flow, neural networks,
biological systems, \emph{etc.}  Graph transformation (or graph
rewriting) is a fundamental concept in concurrency and computational
models, as well in modelling the dynamics of complex system in
general.  From a theoretical point of view, graph rewriting has solid
logic, algebraic and categorical
foundations~\cite{Courcelle90,1997handbook1}, and from a practical
point of view, graph transformations have many applications in
specification, programming, and
simulation~\cite{1997handbook2,1997handbook3}.  Several
graph-transformation languages and tools have been developed, such as
PROGRES~\cite{Schurr97b}, AGG~\cite{ErmelRT97},
Fujaba~\cite{NickelNZ00}, GROOVE~\cite{Rensink03},
GrGen~\cite{GeissBGHS06} and GP~\cite{Plump09}, only to mention a few.

When the graphs are large or growing via transformations, or
when the number of transformation rules is important, being able to
directly interact with the rewriting system becomes crucial to
understand the changes in the graph structure.  From a na\"\i ve point
of view, the output of a graph rewriting system is a dynamic graph: a
sequence of graphs obtained through a series of topological
modifications (addition/deletion of nodes/edges).  However, the study
of a rewriting system is actually much more complex. Reasoning about
the system's properties actually involves testing various rewriting
scenarios, backtracking to a previously computed graph, possibly
updating rules, etc. In this paper, we address these issues. Our main
contribution is a solution to these problems via a strategy-driven
interactive environment for the specification of graph rewriting
systems: PORGY.  

Our work emerged from the necessity to assemble different views on the
rewriting system and relevant interactions.  First of all, an
appropriate formalism and associated structures are needed to
represent and manipulate graph rewriting systems and rewriting
sequences. Our approach is based on the use of port graphs and port
graph rewriting rules~\cite{Andrei08,AndreiK08c}.  We support our
claim on the generality of this concept by using port graph
transformations for modelling biochemical interactions that take part
in the regulation of cell proliferation and transformation.
This case study illustrates the highly dynamic context and
some interesting challenges for graph visualisation provided by
biochemical systems. PORGY provides support for the initial task of
defining a set of graph rewriting rules, and the graph representing
the initial state of the system (the ``initial model'' in PORGY's
terminology), using a visual editor.

Other crucial issues concern when and where rules are applied. To
address this problem, PORGY provides a strategy language to constrain
the rewriting derivations, generalising the control structures used in
PROGRES, GP and rewrite-based programming languages such as Stratego
and ELAN.\@  In particular, the strategy language includes control
structures that facilitate the implementation of graph traversal
algorithms, thanks to the explicit definition of ``positions'' in a
graph, where rules can be applied (we refer the reader
to~\cite{FernandezN10a} for examples of graph programs in PORGY, and
to~\cite{FN10} for the formal semantics of the
strategy language).

Rewriting derivations can also be visualised, and used in an
interactive way, using PORGY's interface.  Designing a graph
transformation system is often a complex task, and the analysis and
debugging of the system involves exploring how rules operate on
graphs, analysing sequences of transformations, backtracking and
changing earlier decisions.  For this purpose, PORGY's visual
environment offers a view on the rewriting history and ways to select
time points in the history where to backtrack.

The organisation of this paper is as follows.  In
Section~\ref{PortGraph}, we recall the concept of port graph and port
graph transformations, and use this formalism to model the scaffold
protein \AKAP\ in the process of mediating a crosstalk between the
\cAMP\ and the \mbox{\Raf/\MEK/\ERK} signalling pathway.  In
Section~\ref{Strategy} we describe PORGY's strategy language.  In
Section~\ref{Visu}, we focus on the visualisation and interaction
features designed to better understand the model and its behaviour.
Section~\ref{sec:related} discusses related work and compares PORGY to
the similar approaches we are aware of.  Finally,
Section~\ref{Conclusion} concludes and describes future work.

\section{Port Graph Rewriting}
\label{PortGraph}

The basic constructs of the PORGY environment are the concept of port
graph and the port graph rewriting relation that we recall in this
section.

Informally, a {\em port graph} is a graph
where nodes have explicit connection points called {\em ports} for the
edges and a {\em p-signature} is a mapping which associates a set of
port names to a node name.

\begin{definition}[P-Signature\cite{Andrei08,AndreiK09}]
  Let $\nabla_{\mathscr{N}}$ be a set of constant node names,
  $\nabla_{\mathscr{P}}$ a set of constant port names, and
  $\X_{\mathscr{P}}$ and $\X_{\mathscr{N}}$ two sets of port name
  variables and node name variables, respectively.  A {\em
    p-signature} is a pair of sets of names \mbox{$\nabla^\X=\langle
    \nabla_{\mathscr{N}}\cup \X_{\mathscr{N}},
    \nabla_{\mathscr{P}}\cup \X_{\mathscr{P}} \rangle$} such that each
  node name $N\in \nabla_{\mathscr{N}}\cup \X_{\mathscr{N}}$ comes
  with a finite set of port names $\Interface(N)\subseteq
  \nabla_{\mathscr{P}}\cup \X_{\mathscr{P}}$. 
\end{definition}


\begin{definition}[Port Graph~\cite{Andrei08,AndreiK09}] 
  Given a fixed p-signature $\nabla^\X$, a {\em labelled port graph over
    $\nabla^\X$} is a tuple $G=\langle V_G,E_G,lv_G, le_G\rangle$ where:
\begin{itemize}
\item $V_G$ is a finite set of nodes;
\item $E_G$ is a finite multiset of edges, \\ $E_G\subseteq\{ \langle
  (v_1,p_1), (v_2,p_2)\rangle \ | \ v_i\in V_G, p_i\in
  \Interface(lv_G(v_i)) \}$;
\item $lv_G:V_G \ra \nabla_{\mathscr{N}}\cup \X_{\mathscr{N}}$ is a
  node-labelling function associating to a node a name and a set of
  ports according to the p-signature;
\item $le_G:E_G\ra (\nabla_{\mathscr{P}}\cup \X_{\mathscr{P}})\times
  (\nabla_{\mathscr{P}}\cup \X_{\mathscr{P}})$ is an edge-labelling
  function associating to an edge the pair of port names where it
  connects to the nodes, i.e. $le_G(\langle (v_1,p_1),
  (v_2,p_2)\rangle) = (p_1, p_2)$.
\end{itemize}
\end{definition}

 \begin{figure}[!t]
    \centering
    \includegraphics[width=0.35\columnwidth]{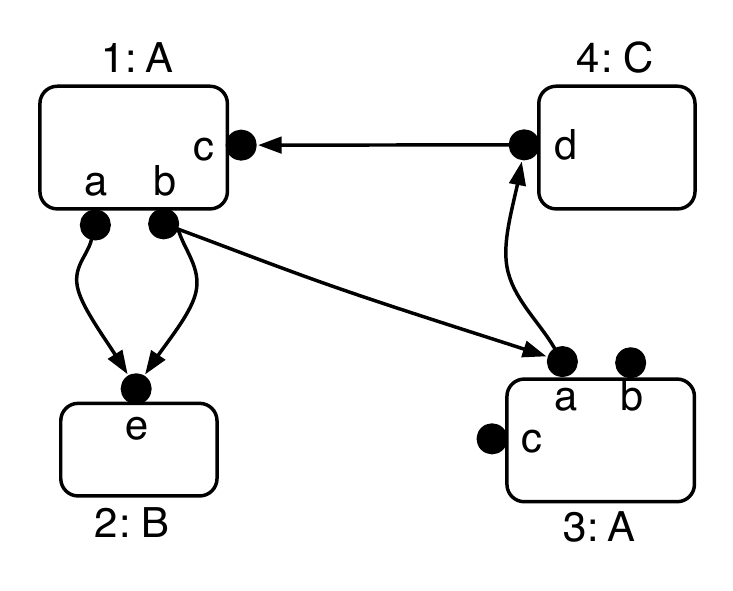}
    \caption{A port graph over the p-signature $\nabla=\langle
      \{A,B,C\}, \{a,b,c,d,e\}$ with $\Interface(A)=\{a,b,c\}$,
      $\Interface(B)=\{e\}$, $\Interface(C)=\{d\}$, and $1,2,3,4$ the
      node identifiers.}\label{fig:portgraph}
  \end{figure}

  A simple example of a port graph is depicted in
  Fig.~\ref{fig:portgraph}. Port graphs were first identified as an
  abstract view of proteins and molecular complexes resulting from the
  protein interactions in a biochemical setting. From a biochemical
  perspective, a protein is characterised by a collection of
  functional domains also called {\em sites}. Two proteins may
  interact by binding on complementary sites. Then a protein with
  binding sites is graphically modelled by a node with ports, and a
  bond between two proteins by an edge in a port graph.  A port can
  also just carry some information from a set of attributes instead of
  being a connection or binding point. For instance, two proteins may
  interact just by changing the attribute information of a site, from
  phosphorylated (P or ``$+$'') to unphosphorylated (U or ``$-$'') and
  vice versa.  This view is at the origin of several formalisms for
  biology, such as the $\kappa$-calculus~\cite{DanosL04} and
  \mbox{BioNetGen}~\cite{FaederBH09}; see
  also~\cite{BournezCCKI03,AndreiIK06}, where graph models have been
  designed to simulate a chemical reactor using rule-based systems and
  strategies.

\begin{figure}[!ht]
    \centering
    \includegraphics[width=0.9\columnwidth]{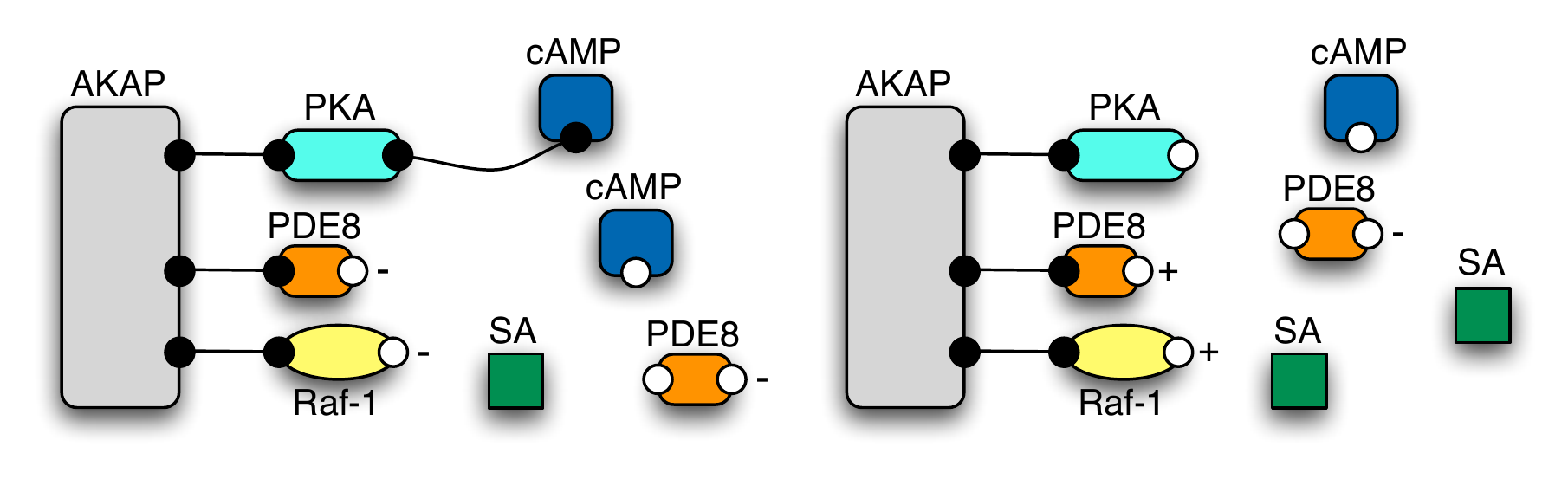}
    \caption{Port graph representation for the chemical species in the
      \AKAP\ model where a port with a plus sign stands for a
      phosphorylated site and a minus sign for unphosphorylated. In
      this graphical representation we omitted the port names as there
      is no risk of confusion.}\label{fig:akap-species}
  \end{figure}

\begin{exmp}[\AKAP\ model: species as port graphs]\label{exmp:akap}
  We illustrate port graphs and port graph rewriting for modelling a
  biochemical network in a simplified model of the scaffold-mediated
  crosstalk between the cyclic adenosine monophosphate (cAMP) and the
  \mbox{Raf/MEK/ERK} pathway~\cite{AndreiC10a}. This interaction has
  an important role in the regulation of cell proliferation,
  transformation and survival.  Let us call this simplified
  biochemical model the {\em \AKAP\ model}. The chemical species
  occurring in the \AKAP\ model are:
  scaffold protein \AKAP\ with three binding sites;
  nucleotide \cAMP\ with one binding site;
  protein \PKA\ with one site for binding to the scaffold and one site
  for binding to \cAMP;
  enzyme \PDE\ with one site for binding to the scaffold and one
  phosphorylation site;
  \Raf\ protein with two sites: one for binding to the scaffold and
  the other for phosphorylation;
  signal protein \SA.
  The \AKAP\ scaffold protein binds the three molecules \PKA, \PDE\
  and \Raf. Although these molecules are not all proteins, we model
  them as nodes with ports in port graphs by abstracting the signal
  transfer as binding actions between their ports.  In
  Fig.~\ref{fig:akap-species} we show a port graph representation of a
  state of the \AKAP\ model.
\end{exmp}

  \begin{definition}[Port graph rewrite rule~\cite{Andrei08,AndreiK09}]\label{pgrule-def}
    A {\em port graph rewrite rule} $L \Ra R$ is a port graph
    consisting of two port graphs $L$ and $R$ over the same
    p-signature and one special node $\Ra$, called {\em arrow node}
    connecting them.  $L$ and $R$ are called the {\em left-} and {\em
      right-}hand side respectively.  The arrow node has the following
    characteristics: for each port $p$ in $L$, to which corresponds a
    non-empty set of ports $\{p_1,\ldots, p_n\}$ in $R$, the arrow
    node has a unique port $r$ and the incident directed edges $(p,r)$
    and $(r,p_i)$, for all $i=1,\ldots,n$; all ports from $L$ that are
    deleted in $R$ are connected to the {\em black hole} port of the
    arrow node.  

    A {\em port graph rewrite system} $\R$ is a finite set of port
    graph rewrite rules.
\end{definition}

Intuitively, the arrow node together with its adjacent edges embed the
correspondence between elements of $L$ and elements of $R$. When the
correspondence between ports in the left- and right-hand side of the
rule is obvious we omit the ports and edges involving the arrow node.

Let $G$ and $H$ be two port graphs defined over the same p-signature.
A {\em port graph morphism} $f:G\ra H$ relates the elements of $G$ to
elements of $H$ by preserving sources and targets of edges, constant
node names and associated port name sets up to a variable renaming.
We say that $G$ and $H$ are {\em homomorphic} when any two ports are
connected in $G$ if and only if their $f$-images are connected in $H$.

We now informally recall the {\em port graph rewriting relation}
from~\cite{Andrei08}.  Let $L\Ra R$ be a port graph rewrite rule and
$G$ a port graph such that there is an injective port graph morphism
$g$ from $L$ to $G$.  By replacing the subgraph $g(L)$ of $G$ by
$g(R)$ and connecting it with the rest of the graph as indicated by
the interface of the rule, we obtain a port graph $G'$ representing a
result of {\em one-step rewriting} of $G$ using the rule $L\Ra R$,
written $G\ra_{L\Ra R}G'$. Several injective morphisms $g$ from $L$ to
$G$ may exist leading to possibly different rewriting results. These
are built as solutions of a {\em matching} problem from $L$ to a
subgraph of $G$.  If there is no such injective morphism, we say that
$G$ is {\em irreducible} with respect to $L\Ra R$.  Given a set $\R$
of rules, a port graph $G$ {\em rewrites} to $G'$, denoted by
$G\ra_{\R}G'$, if there is a port graph rewrite rule $r$ in $\R$ such
that $G\ra_{r}G'$.  This induces a transitive relation on port
graphs. Each {\em rule application} is a rewriting step and a {\em
  derivation} is a sequence of rewriting steps, also called a
computation.  A port graph is in \emph{normal form} if no rule can be
applied on it.  Rewriting is intrinsically non-deterministic since
several subgraphs of a port graph may be rewritten under a set of
rules.

\begin{exmp}[\AKAP\ model: reactions as port graph rewrite
  rules]\label{exemple:akap_contd}
  We consider the following four chemical reactions:
\begin{description}
\item[${\bf (r_1)}$] \cAMP\ activates \PKA\ through binding;
\item[${\bf (r_2)}$] active \PKA\ phosphorylates \PDE\ and \Raf\ on
  the same scaffold and becomes inactive;
\item[${\bf (r_3)}$] phosphorylated \PDE\ degrades free \cAMP\ and
  becomes unphosphorylated as well as \Raf\ at which point \Raf\ sends
  an activation signal \SA;
\item[${\bf (r_4)}$] unphosphorylated \PDE\ degrades free \cAMP\ (we
  represent the binding site for \PDE\ in a shade of grey meaning that
  unphosphorylated \PDE\ has the same behaviour in the presence of a
  \cAMP\ molecule if it is bound or not to the \AKAP\ protein).
\end{description}
These reactions are graphically represented as port graph rewrite
rules in Fig.~\ref{fig:akap-reactions}. A phosphorylation action
activates a site (or port) and we represent this graphically by
changing the attribute ``$-$'' of a site (or port) into ``$+$''; an
unphosphorylation event does the opposite. In
Fig.~\ref{fig:akap-reactions} we only give a schematic representation
of the rule, whereas in Fig.~\ref{fig:akap-reaction4} we detail the
port graph rewrite rule correspoding to reaction {\bf r4}: the arrow
node is included together with two ports for showing correspondence
between the two sites of \PDE\ on both sides of the rule, as well as
the black hole (bh) port which is responsable for deleting the only
one site of \cAMP\ and, in consequence, the entire molecule \cAMP.

\begin{figure}[!t]
  \centering
  \includegraphics[width=1.0\columnwidth]{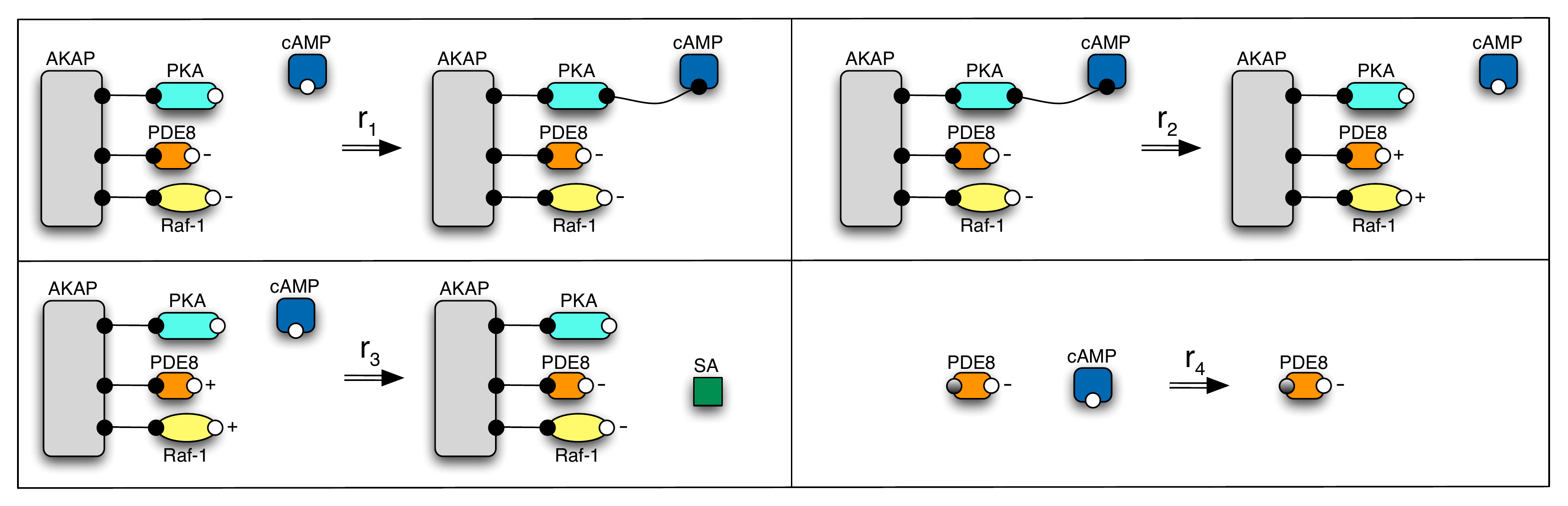}
  \caption{Port graph representation for the
    biochemical reactions in the \AKAP\ model.}
\label{fig:akap-reactions}
\end{figure}

\begin{figure}[!h]
  \centering
  \includegraphics[width=0.6\columnwidth]{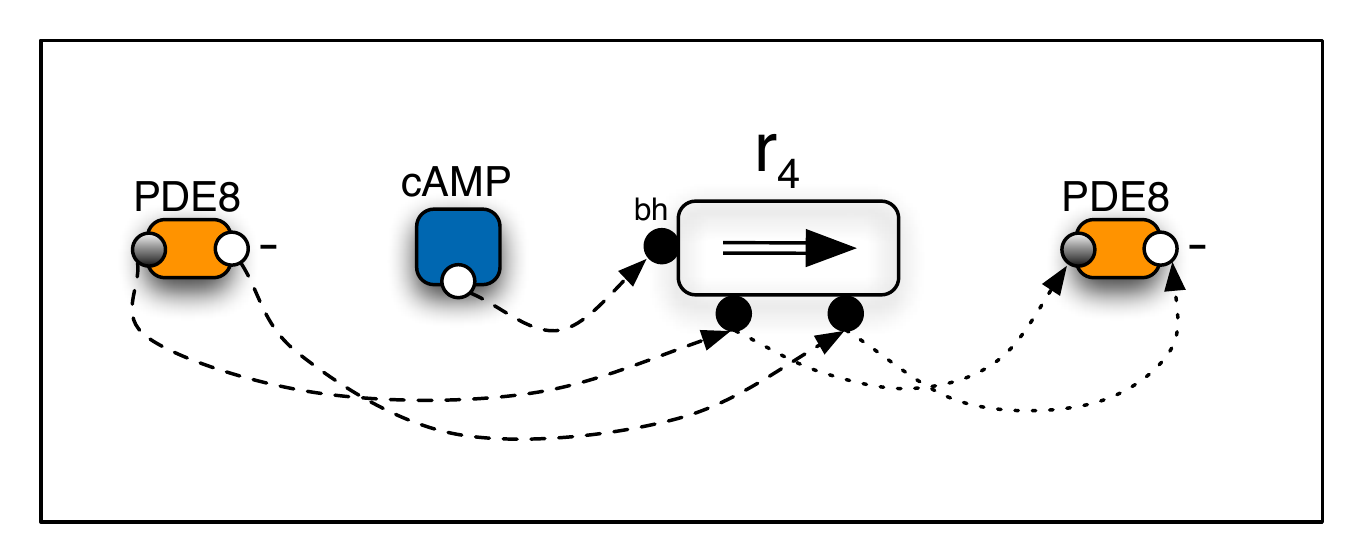}
  \caption{Reaction {\bf r4} as a port graph rule with explicit arrow
    node and {\em black hole (bh)} port.}\label{fig:akap-reaction4}
\end{figure}

An overview of \AKAP's behaviour is as follows. Free \cAMP\ activates
\PKA\ by binding to its free port. When \PKA\ becomes active, it
phosphorylates \PDE\ and \Raf\ if all bound to the same \AKAP\
protein. The amount of \PDE\ may be greater than the amount of \AKAP\
which means that the \PDE\ proteins not on the scaffold protein will
never be phosphorylated.  \PDE\ in either state degrades \cAMP, but
phosphorylated \PDE\ degrades approximately three times more \cAMP\
than unphosphorylated \PDE\ does. This information 
is expressed as the following statement:
{\em rule ${\bf r_3}$ is three times faster than ${\bf r_4}$} in a
stochastic context, or {\em rule ${\bf r_3}$ has 75\% chances of
  occurrence whereas ${\bf r_4}$ only 25\% chances}. As \Raf\ becomes
unphosphorylated, the pathway \mbox{\Raf/\MEK/\ERK} is activated and
the signalling cascade begins; we model this process by the creation
of one signal protein \SA.

Time courses from the laboratory experiments suggest the presence of a
pulsating behaviour in the system. The pulsations ensure that the
state of the \Raf/\MEK/\ERK\ pathway alternates between active and
inactive (note: very long periods of activity or inactivity may
increase the risk of disease). This expected property of the model
translates into alternating short periods of time where the number of
the signal \SA\ increases (active pathway) and times where it remains
constant (inactive pathway).
\end{exmp}

At this point, a comparison of the structure of port graphs with other
graphical formalism needs to be included.  A port graph can be seen as
a multigraph, thus inheriting the theoretical results available for
graph transformations, such as confluence, or parallel independency of
rewriting steps; for more details see~\cite{Andrei08,AndreiK09}. A
graph structure similar to port graphs but with a more restrictive
definition of rewrite rules and rewriting relation can be found in the
context of reserved graph grammars (RGG)~\cite{ZhangZ97} -- a class of
context-sensitive graph grammars well suited for describing and
efficiently implementing \emph{diagrammatic} visual programming
languages. The graphs used in RGGs are based on two-level nodes: a
node is made up of vertices and vertices can be interconnected; one
vertex is distinct for representing the node itself, called
\emph{super vertex}.  The application domain of RGGs imposes the graph
rewrite rules to be locally confluent and vertices in a RGG rule can
be marked which means at matching the incidence degree has to match as
well, ensuring that no dangling edges will occur after a graph
transformation.

\smallskip

In~\cite{Andrei08} a suitable (strategic) rewriting relation and an
abstract calculus have been defined on port graphs. This formalism is
powerful enough to model biochemical applications and generation of
biochemical networks, as well as self-management properties of
autonomic systems~\cite{AndreiK08c,AndreiK09}.

In the study of biochemical networks, the PORGY environment allows
controlled network generation and in silico experiments with different
priorities or probabilities of rule applications. We expect to be able
to answer typical questions asked by biologists, such as, in the
previous example, find the derivations leading to a constant
alternation between active and inactive pathway.

Another main application domain for port graph rewriting and the PORGY
environment, besides the biochemical networks, is the Interaction
Nets~\cite{LafontY:intn,FernandezN10a}.  These applications, amongst others,
motivate our choice of port graphs as a basis for modelling complex
systems and for prototyping their evolution (port graph rewriting
provides executable specifications).  In order to support the various
tasks involved in the study of a port graph rewriting system, we
suggest to combine different points of view on the system:
\begin{itemize}
\item to explore a derivation tree with all possible derivations,
\item to perform on-demand reduction using a strategy language which
  permits to restrict or guide the reductions,
\item to track the reduction throughout the whole tree,
\item to navigate in the tree, for instance, backtracking and
  exploring different branches.
\end{itemize}
These capabilities are developed in the remaining of this paper.

\section{Strategies}
\label{Strategy}

In this section, we present a strategy language that controls the
application of rules from a graph rewriting
system~\cite{FernandezN10a}. This language was designed to cover a
wide range of graphs, but with the main application domains of port
graphs in mind, such as biochemical networks and Interaction Nets.

In the PORGY environment, the user creates a {\em graph-program} from
a strategy over a graph rewriting system, a graph and a position. In
the following, we first review the strategy language and then give
examples of graph-programs.

In term rewriting, by default the rewriting is performed at the {\em
  root} position of a term. Then one may use term traversal strategies
such as top-down, bottom-up, \emph{etc.},\ in order to apply rules or
strategies at particular positions in a term. Graphs generalise the
tree-like structure of terms and lose the notion of root as absolute
position. In order to overcome this, we consider a generalisation of
the notion of position, and define \emph{located graphs}.

\begin{definition}[Located graph]
\label{def:locatedgraph}
Let $G$ be a graph and $P$ a subgraph of $G$ representing the {\em
  position} where a strategy is to be applied. Then the structure
$G[P]$ is called a \emph{located graph}.
\end{definition}

Thus, any subgraph of a graph $G$ can be used as a position $P$ to
apply a rewrite rule.  The notion of position allows us to
\emph{focus} on specific parts of a graph under study, in order to
apply rewrite rules.  Considering a
located graph $G[P]$ to be rewritten, we require that the homomorphic
image of the left-hand side of the rule has a non-empty intersection
with the subgraph $P$. A simple and intuitive example is to define $P$
as consisting of a specific node in $G$, and in this case, only the
rules having this node in the homomorphic image of their left-hand
sides are allowed to be applied. The application of
a strategy on a graph $G$ may change the subgraph $P$ in $G$: our
strategy language includes operators not only to select rules and the
position where the rules are to be applied, but also to change the
focus of the rewriting engine along a derivation.

Figure~\ref{fig:syntax-strategies} shows the grammars $F$, $A$ and $S$
for generating expressions to define  positions, to apply rules
and to define general strategies.  A {\em focusing expression} defines a 
 position subgraph, whereas an {\em application} may change both the graph and the
positions.  A {\em strategy} embeds the previous constructs and
combines them using sequential composition, iteration and conditionals.
In the following we describe informally the semantics of each
operator in the language and give examples.

\begin{figure}[!t]
\centering
\fbox{
\renewcommand{\arraystretch}{1.5}
\begin{tabular}{l}
  Let $G, L, R$ be graphs, $P$ a position, $\rho$ a property, $m,n$
  integers, $p_i \in [0,1]$.
   \\
\begin{tabular}{rrrl}
  {\bf (Focusing)} &  $F$ & $:=$ & ${\tt CrtGraph} \mid {\tt CrtPos} \mid
  {\tt AllSuc} \mid {\tt OneSuc} \mid {\tt NextSuc}$\\
   & & $\mid$ &  $ {\tt Property}(\rho,F)
  \mid  F \cup F \mid
  F \cap F \mid \overline{F} \mid F \setminus F$ \\
  {\bf (Applications)} & $A$ & $:=$ & $\ident \mid \Fail \mid (L \Ra R)_P
  \mid (A\parallel A) \mid (A\interleave A) \mid A^{\parallel(m,n)} $ \\
  {\bf (Strategies)} & $S$ & $:=$ & $  F \mid A \mid S;S  \mid
  S+S \mid {\tt ppick}(S_1/p_1,\ldots ,S_i/p_i) $ \\
  & & $\mid$ & $
  {\tt while}~(S)~{\tt do}(S)~{\tt min}(m)~{\tt max}(n)$
  \\ 
  & & $\mid$ & $ {\tt if}~(S)~{\tt then}(S)~{\tt else}(S) 
  \mid {\tt Empty(F)} \mid \langle S\rangle  \mid {\tt SetPos(}F{\tt )}$ 
\end{tabular}
\end{tabular}
}
\caption{Syntax of the strategy language.}\label{fig:syntax-strategies}
\end{figure}

\smallskip
\noindent 
{\bf Focusing.}  The expressions generated by $F$ allow us to focus on
different parts of the graph during the rewriting process. These
constructs are functions from located graphs to port graphs: an
expression $F$ applies to a located graph $G[P]$ to produce a new
graph $P'$. They are used in strategy expressions to change the
position $P$ where rules apply and to specify different types of graph
traversals.  ${\tt CrtPos}$ returns the position in the current
located graph.  ${\tt AllSuc}$ returns immediate successors of all
nodes in the current position, where an immediate successor of a node
$v$ is a node $u$ with a port connected to a port of $v$. ${\tt
  OneSuc}$ looks for all the immediate successors of all nodes in the
current position and picks one of those non-deterministically.  ${\tt
  NextSuc}$ computes successors of nodes in the current position using
a function that designates a specific port for each node.  ${\tt
  Property}(\rho,Y)$ is a filtering construct, that returns a subgraph
of $G$ containing only the nodes from $Y$ that satisfy the decidable
property $\rho$ ($Y$ would generally be $P$ or $G$, but can be any
graph returned by an expression $F$). $\rho$ typically tests a
property on nodes allowing us, for example, to select the subgraph of
red nodes.  The set theory operators {\em union}, {\em intersection},
{\em complement} and {\em substraction} apply to positions,
considering that those graphs are sets of nodes and edges.

\smallskip
\noindent 
{\bf Applications.}  The application of $\ident$ on a located graph
never fails and leaves the graph unchanged whereas $\Fail$ always
fails (it leaves the graph unchanged and returns failure). $(L \Ra
R)_{Q}$ where $Q$ is a subgraph of $R$, represents the application of the rule $L \Ra R$ at the
current position $P$ in a located graph $G[P]$ 
 where the morphism $g$ is chosen such that $g(L)\cap P$ is not empty;
the current position
becomes $(P \setminus g(L))\cup g(Q)$.
If more than one application is possible, one is non-deterministically selected
from the set of possible results. The other results are then used if there is
a backtrack, that is, if a failure arises.
 $A\parallel A'$ represents
simultaneous application of $A$ and $A'$ on \emph{disjoint} (i.e. not connected) subgraphs of $G$
and returns $\ident$ only if both applications are possible and
$\Fail$ otherwise. $A\interleave A'$ is a weaker version of
$A\parallel A'$ as it returns $\ident$ if at least one application of
$A$ or $A'$ is possible.  $A^{\parallel(m,n)}$ applies $A$
simultaneously a minimum of $m$ and a maximum of $n$ times. If the
minimum is not satisfied then $\Fail$ is returned and $\ident$
otherwise. If $n$ is a negative integer then no maximum is considered.
In the current implementation, these concurrent applications are done
for rules only, \emph{i.e.}\ for an elementary kind of strategies. Extending
these constructions to full strategies needs further exploration.

\smallskip
\noindent
{\bf Strategies.}  
The expression $S;S'$ represents sequential application of $S$
followed by $S'$, and $S+S'$ applies whatever $S$ or $S'$ that returns
$\ident$: if both fail then $\Fail$ is returned and if both are successful
then one of them is picked non-deterministically.
When probabilities $p_1,\ldots,p_n\in [0,1]$ are associated to
strategies $S_1,\ldots,S_n$ such that $p_1+\ldots +p_n = 1$, the
strategy ${\tt ppick}(S_1/p_1,\ldots,S_n/p_n)$ non-deterministically
picks one of the strategies for application, according to the given
probabilities.
For iterations, we have expressions of the form ${\tt while}~(S)~{\tt
  do}(S')~{\tt min}(m)~{\tt max}(n)$ which keep on sequentially
applying $S'$ for as long as the expression $S$ rewrites to $\ident$; if
the minimum of $m$ successful applications of $S'$ is not satisfied
then it returns $\Fail$ or else $\ident$ is returned.  Similar to
$A^{||(m,n)}$, setting $n$ to a negative integer eliminates the
maximum.
The strategy ${\tt if}~(S)~{\tt then}(S')~{\tt else}(S'')$ checks if
the application of $S$ to a located graph $G[P]$ returns $\ident$ in
which case $S'$ is applied to $G[P]$ otherwise $S''$ is applied. $S$
is only tested on $G[P]$ and does not actually change the located
graph.  ${\tt Empty}$ returns $\ident$ if the current position is
empty and $\Fail$ otherwise.  This can be used for instance inside
the condition of an ${\tt if}$ or ${\tt while}$, to check if the
application of the strategy makes the current position empty or not,
instead of checking if the strategy itself can be applied.  The
strategy $\langle S\rangle$ applies $S$ and considers $S$ as one
atomic rewriting step in the derivation tree. This is useful to
abstract several reduction steps as one for visualisation purposes.
${\tt SetPos}(P)$ changes the current
position to a new position $P$.

\begin{definition}[Graph-program]
\label{def:graphprog}
A \emph{graph-program} is a pair $\sema{S_{\cal R}, G[P]}$ where
$S_{\cal R}$ is a strategy expression built over a graph rewriting
system $\cal R$ and $G[P]$ a located graph.  The result of the
execution of a terminating graph-program is another graph-program of
the form $\sema{\ident, G'[P']}$ or $\sema{\Fail, G'[P']}$ such that
$G'[P']$ has been obtained by the application of $S_{\cal R}$ on
$G[P]$.

\end{definition}

The semantics of the strategy constructors defined by the grammars in
Fig.~\ref{fig:syntax-strategies} has been formally defined
in~\cite{FN10} using rewrite rules that reduce a graph-program
$\sema{S, G[P]}$.

The notion of graph-program defined above is very general, and the
language allows programmers to define high-level algorithms in a
variety of application domains.  For examples of programs developed
using this language, we refer the reader
to~\cite{FernandezN10a}. Below we describe an example that exploits
the features of the language to simulate the behaviour of a
biochemical system.





\begin{exmp}[Strategy for the \AKAP\ model]\label{exmp:strat}
  In the AKAP model introduced in Sect.~\ref{PortGraph} the initial
  port graph $G_0$ to be rewritten consists of: 200 unbound \cAMP\
  molecules; 10 structures built upon an \AKAP\ protein binding an
  inactive \PKA, an unphosphorylated \PDE\ and an unphosphorylated
  \Raf; and 3 unphosphorylated \PDE\ proteins not bound to an \AKAP\
  scaffold protein.

  From the lab experiments, the biologists concluded that
  phosphorylated \PDE\ degrades three times more \cAMP\ than the
  unphosphorylated \PDE, in other words reaction rule {\bf $r_3$} is
  three times faster than reaction rule {\bf $r_4$} in a stochastic
  setting. We model this behaviour in a probabilistic setting via the
  strategy construct ${\tt ppick}$ which applies the rule {\bf $r_3$}
  with probability 0.75 and the rule {\bf $r_4$} with probability
  0.25. We remark that {\bf $r_1$} and {\bf $r_4$} have a critical
  pair, therefore we use the strategy ${\tt ppick}$ with probability
  0.5 for each of the rules; the same reasoning goes for {\bf $r_2$}
  and {\bf $r_4$} in the presence of a free \cAMP\ molecule. In order
  to generate the biochemical network, we repeat the applications of
  all rules according to their application probabilities until we
  reach a normal form. Therefore the strategy for the \AKAP\ model has
  the following form:
  $$S_{\AKAP}\ =\ {\tt repeat}_*({\tt ppick}(r1/0.5,r4/0.5),\
  {\tt ppick}(r2/0.5,r4/0.5),\ {\tt ppick}(r3/0.75,r4/0.25))$$
  where ${\tt repeat}_* (S)$ is syntactic sugar for ${\tt
    while}(S)~{\tt do}(S)~{\tt min}(1)~{\tt max}(-1)$. We show the
  result of the application of $S_{\AKAP}$ on $G_0$ in
  Sect.~\ref{Visu} as well as a procedure to analyse the results by
  counting the number of signal proteins \SA\ during the execution.
\end{exmp}

\section{Working with the PORGY platform}
\label{Visu}

\begin{figure}[ht]
\centering
\includegraphics[width=1\columnwidth, keepaspectratio]{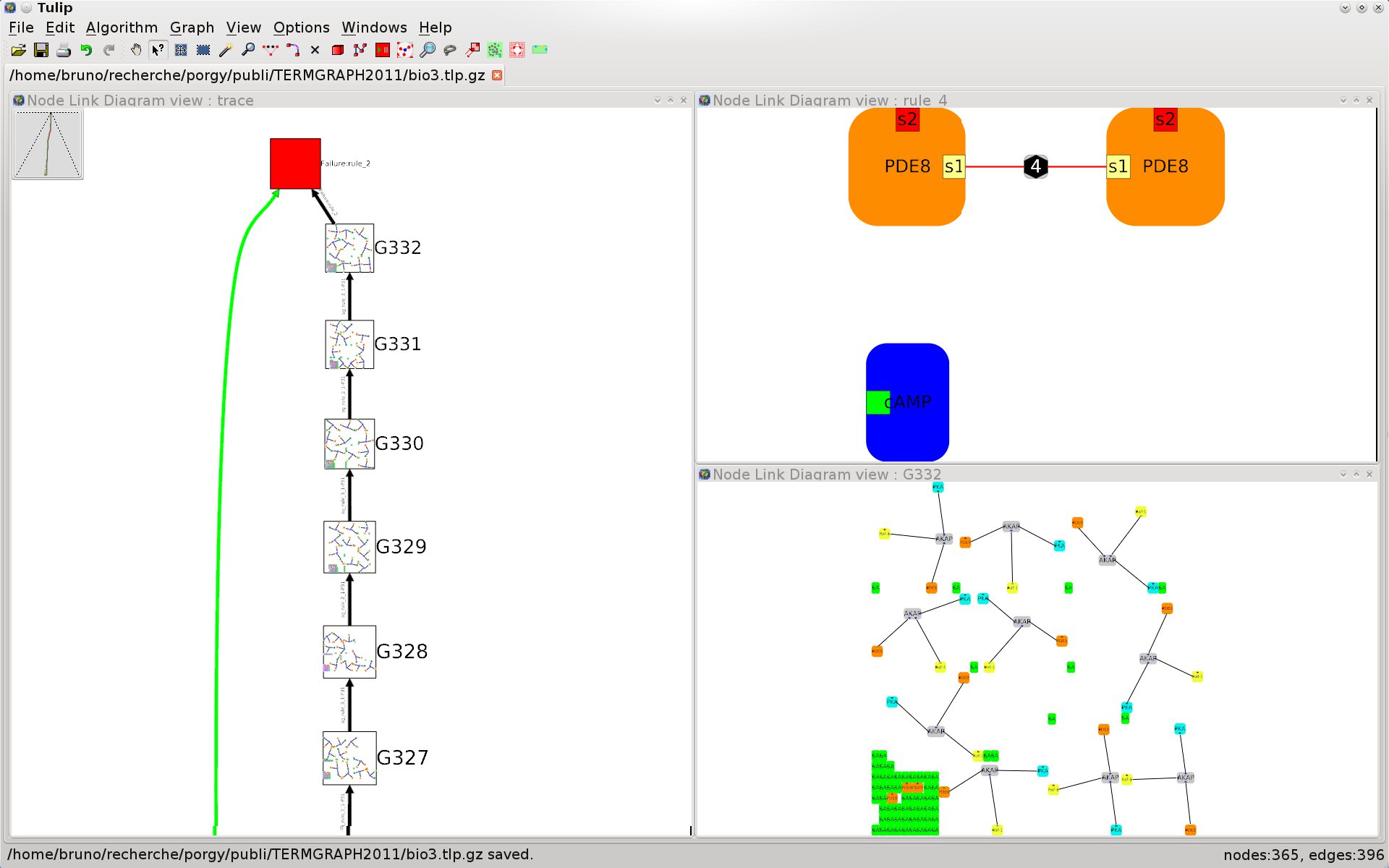}
\caption{Overview of the PORGY environment. The graph at the bottom
  right ($G332$) corresponds to the normal form of $G_{0}$ after application of the
  strategy $S_\AKAP$ (see Example~\ref{exmp:strat}). The
  left panel shows part of a derivation tree, with nodes of the tree containing
  the last intermediate states of $G$. The top right panel corresponds to the
  rule ${\bf r_4}$.}\label{fig:systreecr}
\end{figure}

One of the goals of PORGY is to allow the user to interact and
experiment with a port graph rewriting system in a visual and
interactive way. Ideally, a visual environment should offer
different views on each component of the rewriting system: the
current graph being rewritten, the derivation tree and the rules as
shows the overview (for the \AKAP\ model) in
Figure~\ref{fig:systreecr}. The application of rules is performed
based on an ad hoc matching algorithm finding instances
of left-hand sides of rules in a port graph. Our matching algorithm
is based on the work of Ullman~\cite{ullmann:76} and Cordella \emph{et
  al.}~\cite{cordella:04}. A detailed discussion
of our matching algorithm is out of scope here.

As the figure shows, the normal form of $G_0$ is found, after the
strategy $S_\AKAP$ has been successfully applied with $S_\AKAP$ and
$G_0$ defined in Example~\ref{exmp:strat}. Due to its design, the
strategy terminates on a failure because no more rules apply. The
failure is made explicit by showing a red node as part of the trace
tree (the graph shown on the left pane). We make use of a quotient graph
to embed the states of $G$ into nodes of the trace tree (nodes G327 to G332).
Thus, when scanning graphs along a branch, one can read the
evolution of the graph being rewritten and explore the effect and properties of a strategy.
A local and more detailed view allows a closer examination of
a particular state of $G$ (graph G332 in the lower right pane).

\begin{figure}[ht]
\centering
\includegraphics[width=1\columnwidth, keepaspectratio]{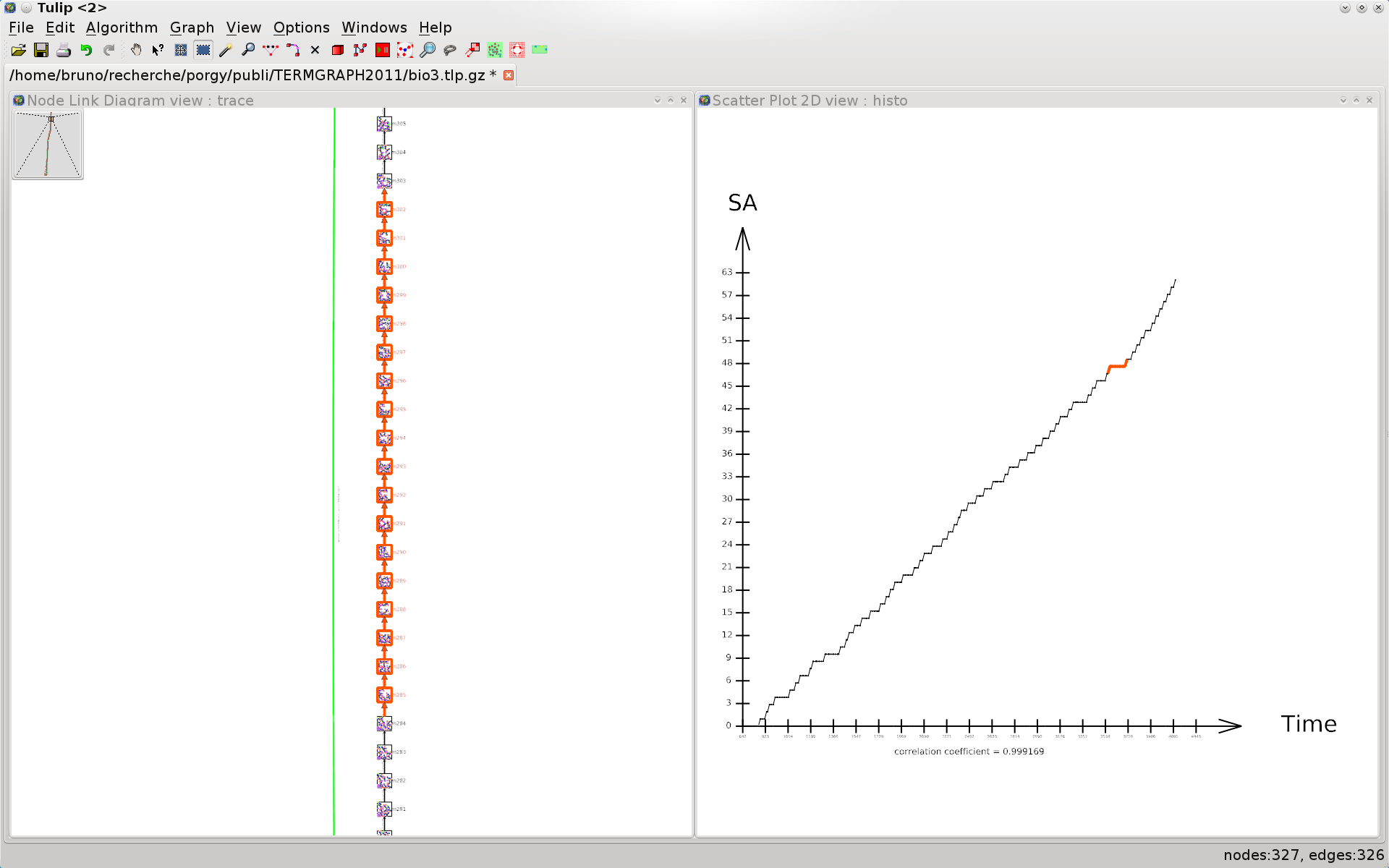}
\caption{Evolution of the number of signal protein \SA\ when
  applying the strategy presented in Example~\ref{exmp:strat}
  -- until a graph in normal form is found.}
\label{fig:sa_evolution}
\end{figure}

Typically, the user may be interested in plotting the evolution of a parameter computed out
of each intermediary state. For example, going back to the \AKAP\ model (Example~\ref{exemple:akap_contd})
the behaviour of the \SA\ protein, as predicted by the biologists, can be examined by
plotting the curve of the evolution of the number of \SA\ protein throughout the rewriting
process (Figure~\ref{fig:sa_evolution}).

\begin{figure}[ht]
\centering
\includegraphics[width=1\columnwidth, keepaspectratio]{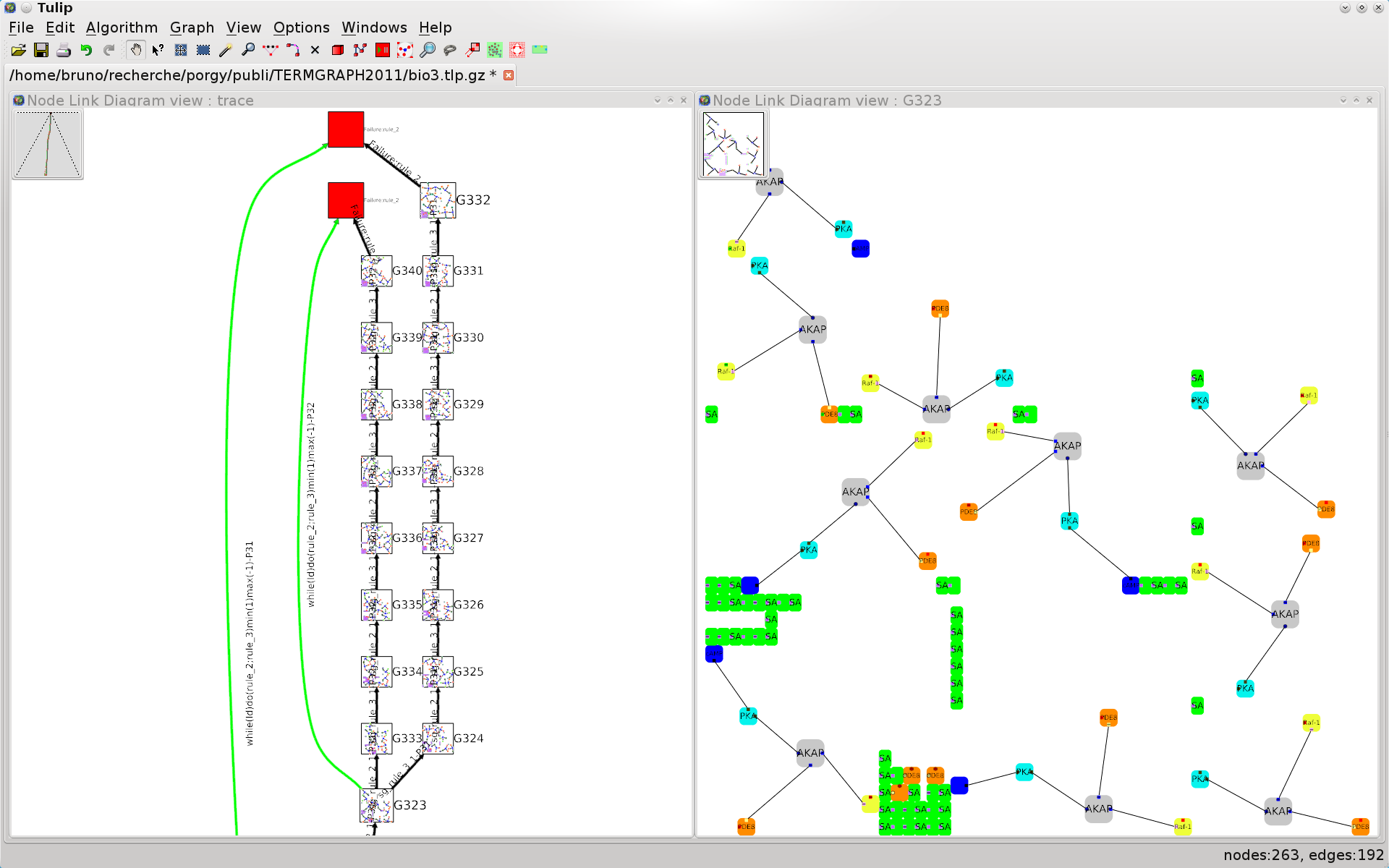}
\caption{The green edges in the derivation tree (the graph on the left and the green edges are the edges on the left of this graph) represent the start and end of previously applied strategies. The other
  edges of the derivation tree represent the application of a single rewriting rule between two
  states of $G$.}\label{fig:strat_visu}
\end{figure}

The interactive features of PORGY simplify the study of the rewriting system
partly due to a synchronisation between the different views. For example,
selecting points on the plot view (Fig.~\ref{fig:sa_evolution})
triggers the selection of the corresponding nodes in the trace tree.
Such a mechanism obviously helps to track properties of the output graph
along the rewriting process. Alternate strategies can be applied from any current or past state
(see Fig.~\ref{fig:strat_visu}).

The system we are currently developing is built on top of the Graph Visualisation framework
Tulip~\cite{tulip2003}\footnote{See also
  \url{http://www.tulip-software.org}.}.
Specific plugins have been designed and developed to implement the various
interaction required to simulate and study port graph rewriting systems.
Because Tulip does not handle port graph directly, each port node is built from several standard nodes,
while implementation details exploiting Tulip's built-in graphical features
have been kept hidden to the user.

\section{Related Work}
\label{sec:related}

There are several tools available for editing graphs, of which some
allow users to model graph transformations. In this section we review
the ones that are similar in scope with PORGY.

{\bf GROOVE}~\cite{Rensink03} is centered around the use of simple
graphs for modelling the design-time, compile-time, and run-time
structure of object-oriented systems.  The GROOVE tool set includes an
editor for creating graph production rules, a simulator for visually
computing the graph transformations induced by a set of graph
production rules, a generator for automatically exploring state
spaces, a model checker for analysing the resulting graph
transformation systems and an imaging tool for converting graphs to
images.  Visualisation is not its main objective, and after each
rewrite step the user must update the layout of the graph by hand. The
model transformations and the operational semantics are based on graph
transformations. GROOVE permits to control the application of rules,
via a control language with sequence, loop, random choice, conditional
and simple (non recursive) function calls.  These are similar to
PORGY's constructs, but the main difference is that GROOVE's language
does not include the notion of position, it is not possible to specify
a position for the application of rules within the language.  Tracing
is possible through state space exploration.

{\bf Fujaba}~\cite{NickelNZ00} Tool Suite is an Open Source CASE tool
providing developers with support for model-based software engineering
and re-engineering.  It combines UML class diagrams, UML activity
diagrams, and a graph transformation language and offers a formal,
visual specification language that can be used to completely specify
the structure and behaviour of a software system under development.
Graphs and rules are used to generate Java code. Fujaba has a basic
strategy language, including conditionals, sequence and method
calls. There is no parallelism, and again one of the main differences
with PORGY is that Fujaba does not include a notion of position to
guide the rule application.

{\bf AGG}~\cite{ErmelRT97} is a rule-based visual language supporting
an algebraic approach to graph transformation, and is aimed at
specifying and implementing applications with complex graph-structured
data. AGG may be used as a general purpose graph transformation engine
in high-level JAVA applications employing graph transformation
methods.  Rule application can be controlled by defining layers and
then iterating through and across these layers. Again, the position
can not be specified and there is no control on the search for
redexes.

{\bf PROGRES}~\cite{Schurr97b} project works on the theoretical
foundations as well as the practical implementation of an executable
specification language based on graph rewriting systems (graph
grammars). It combines EER-like and UML-like class diagrams for the
definition of complex object structures with graph rewrite rules for
the definition of operations on these structures.  PROGRES allows
users to define the way rules are applied (it includes
non-deterministic constructs, sequence, conditional and looping) but
it does not allow users to specify the position where the rule is
applied. It is a very expressive language and also includes a tracing
functionality through backtracking.

{\bf GrGen.NET}~\cite{GeissBGHS06} is a programming tool for graph
transformation designed to ease the transformation of complex graph
structured data as required in model transformation, computer
linguistics, or modern compiler construction, for example. It is
comparable to other programming tools like parser generators which
ease the task of formal language recognition. GrGen.Net has a rule
application language with constructs for sequential, logical and
iterative application. 

{\bf GP}~\cite{Plump09} is a rule-based, non-deterministic programming
language. Programs are defined by sets of graph rewriting rules and a
textual expression that describes the way in which rules should be
applied to a given graph.  The simplest expression is a set of rules,
and this means that any of the rules can be applied to rewrite the
graph. The language has three main control constructs: sequence,
repetition and conditional (if-then-else), and it has been shown to be
complete.  It uses a built-in Prolog-like backtracking technique: if
at some point no rule can be applied, it backtracks to the nearest
point where there was a choice of redex (users cannot easily handle
the derivation tree or change the backtracking algorithm).

{\bf GReAT}~\cite{BalasubramanianNBK06} is a tool for building model
transformation tools. First, one has to specify the metamodels of the
input and target models (using UML style class diagrams) and give
rules to specify the transformation.  Rules are pairs of typed,
attributed graphs. Then, the pattern-matching algorithm always starts
from specific nodes called ``pivot nodes''. Rule execution is
sequenced and there are conditional and looping structures.

\smallskip

PORGY and its strategy language allow a higher expressive power
with its focus on position. Strategies are not limited to picking
random applications but can travel through the graph in a dynamic and
strategic manner to apply rules and sub-strategies.  PORGY has also a
strong focus on visualisation and scale, thanks to the TULIP back-end
which can handle large graphs with millions of elements and comes with
powerful visualisation and interaction features. Some of Tulip's built-in
functionalities, such as selecting some nodes in a visualisation for highlighting the equivalent
nodes in the trace tree, give the user an immediate visual
feedback (Fig~\ref{fig:sa_evolution}).

\section{Conclusion and Future Work}
\label{Conclusion}

The PORGY environment provides an interactive visual environment for
graph transformations. In this paper we presented the main concepts
underlying PORGY: port graph rewriting and strategies for graph
rewriting.

The PORGY environment is yet under development.  
A first implementation of the strategy language is available but needs 
further improvement.
From the
visualisation point of view, we are working on enhancing the
algorithms for drawing rules and models.

Verification and debugging tools for avoiding conflicting rules or non
termination for instance are also planned in the future.  Moreover we
will address other application domains: for instance linguistics
analysis~\cite{FoxLC} shares some of the features of biological
networks and we expect to be able to handle linguistic models in PORGY
without much difficulty. PORGY already provides information about the
number of matching solutions for the application of a rule. Based on
this information we plan to extend a port graph rewriting system with
a stochastic semantics~\cite{KrivineMT08} which is very useful for
developing biochemical models.

\bibliographystyle{eptcs}
\bibliography{bib}

\end{document}